\documentclass[aps,reprint,superscriptaddress,nofootinbib,amsmath,amssymb,%
draft]{revtex4-2}
\usepackage{mathrsfs}
\usepackage{xcolor}

\begin{document}

\title{Non-Noetherian conformal scalar fields}

\author{Eloy Ay\'on-Beato}
\email{ayon-beato-at-fis.cinvestav.mx} \affiliation{Departamento de
F\'{\i}sica, Cinvestav, Apartado Postal 14740, 07360 CDMX,
M\'exico.}
\author{Mokhtar Hassaine}
\email{hassaine-at-inst-mat.utalca.cl} \affiliation{Instituto de
Matem\'atica, Universidad de Talca, Casilla 747, Talca, Chile.}

\begin{abstract}
Recently, an extension of the standard four-dimensional scalar conformal
action, yielding a second-order field equation that remains conformally
invariant, was proposed. In spite of this, the corresponding action is not
invariant under conformal transformations and this motivates us to define
the notion of non-Noetherian conformal scalar field. In this article, we go
further by determining the most general action in four dimensions that
gives rise to a non-Noetherian conformal scalar field satisfying a
second-order equation. This task is achieved by using the solution to the
inverse problem of the calculus of variations. Surprisingly enough, the
standard equation is shown to be extended by a non-Noetherian conformal
piece involving a nonminimal coupling with a very particular combination of
squared curvature terms, which is none other than the one defining the
so-called Critical Gravity. We also prove that the most general
second-order Euler-Lagrange equation for a conformal scalar field involves
additional Noetherian conformal nonminimal couplings defined by an
arbitrary function of the Weyl tensor. The recently proposed non-Noetherian
conformal extension is recovered as a particular example of this function.
\end{abstract}

\maketitle

The prominence of conformal symmetry in Physics and Mathematics is beyond any
doubt. Concerning Physics, understanding the physical properties that are
locally scale independent is one of the fundamental questions that conformal
transformations attempt to explain. Therefore, these transformations are
exploration tools in various fields ranging from gravitational physics to
high energy physics, including condensed matter physics, and not forgetting
statistical physics. In the latter, for example, one of the greatest
challenges from a theoretical and practical point of view is to understand
under which conditions conformal invariance is expected in critical
phenomena. Among other things, this is relevant to explain the universality
of critical exponents, which in general cannot be determined in a precise way
from the scaling hypothesis alone (without appealing to conformal
invariance). Conversely, critical exponents can be fully determined from the
so-called conformal bootstrap, see e.g.~\cite{Poland:2018epd}. Actually, the
relevance of conformal symmetry in general field theories has become
extremely appreciable from the fact that any QFT appears to be a RG flow from
UV to IR between field theories invariant under the conformal group of flat
spacetime, and not only under the Poincar\'e one \cite{Polchinski:1992ed}.
More recently, it has been shown that some of these CFTs are even amenable to
holographic descriptions allowing to study their strongly correlated sector
\cite{Maldacena:1997re}, which brings hope on how to describe some of the SM
processes inaccessible to the standard perturbative approach, as for example
the confinement. Conformal invariance also naturally emerges in Relativity,
since one important property preserved by conformal transformations is the
causal structure of spacetime. Causality is determined by light rays
described with the Maxwell theory, which became the first known example of
conformal field theory realized in Nature. Historically, the idea that at
high energy Nature can be described by a theory without scale and where the
scale only emerges at low energy finds its origin in the seminal work of Weyl
\cite{Weyl:1918ib}. His main goal was in fact to propose a unified theory of
gravity and electromagnetism based on a generalization of Riemannian
geometry. Although Weyl's theory turned out to be physically unsatisfactory,
it has shed light on gauge theories and what is now known as conformal
gravity; an alternative to GR in the hope of being able to explain the dark
sector of the Universe \cite{Mannheim:2005bfa}. Regarding applications of
conformal symmetry in Mathematics, they are also numerous but we will limit
ourselves to the case of the famous Yamabe problem \cite{Lee:1987}. Indeed,
it is one of the most important problems in conformal geometry and its
solution has marked an important step in the development of the theory of
nonlinear partial differential equations. Mathematically speaking, for a
given Riemannian manifold the Yamabe problem consists of finding a conformal
metric having a constant scalar curvature. The nonlinear eigenvalue problem
satisfied by the involved conformal factor is connected to gravitational
physics, since it is nothing but the standard conformally invariant scalar
field equation on curved spacetimes. This equation arises from the variation
of the Klein-Gordon action supplemented with a precise nonminimal coupling
between the scalar field and the scalar curvature
\cite{Gursey:1963,Penrose:1965,Callan:1970ze}. Remarkably, the related
self-gravitating problem gave rise to the first example of an asymptotically
flat black hole dressed with a conformal scalar field
\cite{BBM,Bekenstein:1974sf}. It should be stressed that the conformal
invariance of the scalar field plays an important role in constructing this
solution. Later on, its extension in presence of a cosmological constant was
achieved by adding a self-interacting potential that preserves the conformal
invariance of the scalar field \cite{Martinez:2002ru}. From
the discussion above, we consider that conformal symmetry is
fundamental to broad our knowledge beyond our current understanding.

Of particular importance for what follows, we would like to underline that
any symmetry of an equation resulting from a principle of action is in
general not verified at the level of the action. In this precise case, it
seems to us natural to call it a \emph{non-Noetherian symmetry}. Having
clearly specified the terminology, we can safely say that this article will
primarily focus on \emph{non-Noetherian conformal scalar fields}. To our
knowledge, the first such example was pointed out by Jackiw in two dimensions
\cite{Jackiw:2005su}. More recently, a four-dimensional example was provided
by Fernandes where the non-Noetherian conformal symmetry was achieved thanks
to a nonminimal coupling of the scalar field to the Gauss-Bonnet density
\cite{Fernandes:2021dsb}. In this case, the conformally invariant equation
can be written in a conveniently compact form as
\begin{equation}\label{eq:Fernandes}
\Box\Phi-\frac16R\Phi-4\lambda\Phi^3-\alpha\Phi^3\tilde{\mathscr{G}}=0,
\end{equation}
where $\lambda$ and $\alpha$ are two coupling constants, $\Box$ and $R$ stand
respectively for the d'Alembertian operator and the scalar curvature of the
metric $g_{\mu\nu}$. Meanwhile, $\tilde{\mathscr{G}}=
\tilde{R}^2-4\tilde{R}_{\alpha\beta}\tilde{R}^{\alpha\beta}
+\tilde{R}_{\alpha\beta\mu\nu}\tilde{R}^{\alpha\beta\mu\nu}$ is the
Gauss-Bonnet density constructed out of the curvature components of an
auxiliary metric defined by $\tilde{g}_{\mu\nu}=\Phi^2g_{\mu\nu}$. The great
novelty of this equation is precisely the $\alpha$-piece, since the remaining
terms are nothing but those associated to the standard Noetherian conformal
scalar field equation already mentioned
\cite{Gursey:1963,Penrose:1965,Callan:1970ze}. The conformal invariance of
Eq.~\eqref{eq:Fernandes} can be easily checked by noticing that the auxiliary
metric, and consequently its Gauss-Bonnet density $\tilde{\mathscr{G}}$, are
manifestly invariant under the conformal transformations defined by
\begin{equation}\label{eq:ConfTransf}
g_{\mu\nu} \mapsto \Omega(x)^2 g_{\mu\nu},\qquad \Phi \mapsto
\Omega(x)^{-1}\Phi,
\end{equation}
which are exactly those that leave also invariant the standard conformal
scalar field equation when $\alpha=0$. This explains the convenience of
writing Eq.~\eqref{eq:Fernandes} using a single term as $\tilde{\mathscr{G}}$
instead of displaying a rather long formula with respect to the metric
$g_{\mu\nu}$. In fact, the full equation \eqref{eq:Fernandes} can be
re-expressed in terms of scalar quantities built out of the auxiliary metric
$\tilde{g}_{\mu\nu}$ as
\begin{equation}\label{eq:Fernandestilde}
-\frac16\tilde{R}-4\lambda-\alpha\tilde{\mathscr{G}}=0.
\end{equation}
Although, it is clear that any scalar quantity constructed using the
auxiliary metric $\tilde{g}_{\mu\nu}$ and its curvature can be added to the
above equation without spoiling the conformal invariance nor the second
order, the real challenge is to determine those quantities which come from
the variation of an action principle. As we will see later, this last
requirement will considerably restrict the possibilities. The involved action
for Eq.~\eqref{eq:Fernandes} was found by Fernandes in
\cite{Fernandes:2021dsb}, and is given by {\small
\begin{align}
S_{\lambda,\alpha}[\Phi,g]=&\int\!d^4x\sqrt{-g}\biggl[
- \frac12(\nabla\Phi)^2 - \frac1{12}R\Phi^2 - \lambda\Phi^4 \nonumber\\
&\qquad\qquad\quad\!\enspace- \alpha\biggl(\ln(\Phi)\mathscr{G}
- \frac4{\Phi^2}G^{\mu\nu}\nabla_\mu\Phi\nabla_\nu\Phi \nonumber\\
&\qquad\qquad\quad\!\enspace- \frac4{\Phi^3}(\nabla\Phi)^2\Box\Phi +
\frac2{\Phi^4}(\nabla\Phi)^4\!\biggr)\biggr],
\label{eq:actionFernandes}
\end{align}}%
where $G^{\mu\nu}$ is the Einstein tensor and $\mathscr{G}$ the Gauss-Bonnet
density. For $\alpha=0$, one recovers the well-known action for a standard
conformal scalar field whose invariance under the conformal transformations
\eqref{eq:ConfTransf} is ensured up to boundary terms. This is no longer the
case for the $\alpha$-contribution of the action which, as was carefully
elucidated in \cite{Fernandes:2022zrq}, explicitly breaks the conformal
symmetry. Consequently, the conformal invariance of \eqref{eq:Fernandes} has
not a Noetherian origin as already mentioned. In addition to its
non-Noetherian conformal symmetry, this theory has some interesting
curiosities; whether it is the fact that the $\alpha$-extension is the same
allowing a consistent four-dimensional Gauss-Bonnet gravity (see
\cite{Fernandes:2022zrq} for a recent review) or by the aspect that the full
action is a subclass of the Horndeski theories \cite{Horndeski:1974wa}
yielding second-order field equations not only for the scalar field $\Phi$
but also for the metric $g_{\mu\nu}$.

From these results, it is legitimate to go further and to ask
\emph{what is the most general action in four dimensions giving a
conformally invariant second-order equation for the scalar field?}
This is the main question we address in this letter. In order to
attack this problem, it is more convenient to work in the
exponential frame for the scalar field as defined by $\Phi=e^\phi$.
In doing so, the conformal transformations \eqref{eq:ConfTransf} are
equivalently reformulated as
\begin{equation}\label{eq:ConfTransfExp}
g_{\mu\nu} \mapsto e^{2\omega(x)} g_{\mu\nu},\qquad \phi \mapsto
\phi-\omega(x),
\end{equation}
and their infinitesimal versions read
\begin{equation}\label{eq:ConfTransfInf}
\delta_\omega g_{\mu\nu} = 2\omega g_{\mu\nu},\qquad
\delta_\omega\phi = -\omega,
\end{equation}
while the auxiliary metric is now defined by
$\tilde{g}_{\mu\nu}=e^{2\phi}g_{\mu\nu}$.
The starting point is to specify what would be the generic form of a
second-order equation invariant under these transformations. First
of all, this equation must be a function of $\phi$ and its
derivatives up to second order, together with the metric tensor
$g_{\mu\nu}$ and its associated curvature. Working with the
auxiliary metric $\tilde{g}_{\mu\nu}$, this equation can be
schematically written as
\begin{equation}\label{eq:2do}
E(\phi,\tilde{\nabla}_\mu\phi,\tilde{\nabla}_{\mu}\tilde{\nabla}_{\nu}\phi,
\tilde{g}_{\mu\nu},\tilde{R}^\alpha_{~\beta\mu\nu})=0.
\end{equation}
Requiring now this equation to be conformally invariant imposes that
\begin{equation}\label{eq:infinCI}
-\delta_\omega E = \frac{\partial E}{\partial\phi}\omega
+\frac{\partial
E}{\partial\tilde{\nabla}_\mu\phi}\tilde{\nabla}_\mu\omega
+\frac{\partial
E}{\partial\tilde{\nabla}_\mu\tilde{\nabla}_{\nu}\phi}
\tilde{\nabla}_\mu\tilde{\nabla}_{\nu}\omega=0,
\end{equation}
since by virtue of (\ref{eq:ConfTransfInf}),
$\delta_\omega\tilde{g}_{\mu\nu}=0$. Moreover, since the condition
(\ref{eq:infinCI}) must hold for any conformal factor $\omega$, this
requires the expression $E$ to be independent of the scalar field
and its covariant derivatives with respect to the auxiliary metric,
namely
\begin{equation}\label{eq:2doCI}
E(\tilde{g}_{\mu\nu},\tilde{R}^\alpha_{~\beta\mu\nu})=0.
\end{equation}
This first result also highlights that representations in the form
of \eqref{eq:Fernandestilde} and their generalizations are not only
useful for making conformal invariance manifest, but rather that any
conformally invariant second-order scalar field equation can always
be written in terms of the geometric invariants built out of the
auxiliary metric.  A similar conclusion was also achieved in
\cite{Fernandes:2021dsb} using some clever conformal frame arguments
of Ref.~\cite{Padilla:2013jza}.

As already mentioned, the most difficult part is to single out the
subfamilies of the infinite set of second-order conformally
invariant equations \eqref{eq:2doCI} that arise from an action
principle. Fortunately, this leads to an old problem already solved,
known as \emph{the inverse problem of the calculus of variations}
\cite{Olver:1986}, which consists in asking more generally if a set
of equations can be seen as the Euler-Lagrange equations of some
action. Historically, the first person to consider this problem was
Helmholtz \cite{Helmholtz:1887}, and the conditions ensuring the
existence of a variational principle associated to the equations are
now known as the Helmholtz conditions. But \emph{Quae sunt Caesaris,
Caesari}, and the general solution was in fact discovered by
Volterra \cite{Volterra:1913}. In order to impose the generalized
Helmholtz conditions found by Volterra in our specific problem, and
for the purpose of being self-contained, a couple of definitions
will be given (for more details, see the excellent book of Olver
\cite{Olver:1986}).

First, let $P[u]=P(x,u^{(n)})$ be a differential function, i.e.\ that depends
on the point $x$, the function $u$ together with its derivatives $u^{(n)}$.
Consider now its variation under a one-parameter family of functions. After
interchanging the variation with derivatives (and without integrating by
parts) we end with a differential operator acting on an arbitrary variation
$\delta u$ called the \emph{Fr\'echet derivative} of $P$ (see \cite[p.\
307]{Olver:1986})
\begin{equation}\label{eq:Frechet}
\delta P=\frac{\mathrm{d}}{\mathrm{d}\varepsilon} P[u+\varepsilon
\delta u]\Bigr|_{\varepsilon=0}\equiv\mathrm{D}_P(\delta u).
\end{equation}
Our characterization of the Fr\'echet derivative in terms of variations is
instrumental for its straightforward calculation. In the studied problem, we
consider the second-order conformally invariant pseudoscalar defined by
\begin{equation}\label{eq:2doCIps}
\mathscr{E}=\sqrt{-\tilde{g}}\,
E(\tilde{g}_{\mu\nu},\tilde{R}^\alpha_{~\beta\mu\nu}),
\end{equation}
which is the natural quantity that could be derived from a covariant
action. Here, the role of the dependent function $u$ is played by
the scalar field $\phi$, and hence the Fr\'echet derivative of
$\mathscr{E}$ can be calculated from
\begin{equation}\label{eq:deltaE}
\mathrm{D}_\mathscr{E}(\delta\phi)=\delta_\phi\mathscr{E},
\end{equation}
resulting in the following operator
\begin{equation}\label{eq:FrechetE}
\mathrm{D}_\mathscr{E}=4\sqrt{-\tilde{g}}\left(E-\frac12\tilde{g}^{\mu\nu}
\frac{\partial E}{\partial\tilde{g}^{\mu\nu}}
-\tilde{P}^{\mu\nu}\tilde{\nabla}_{\mu}\tilde{\nabla}_{\nu}\right).
\end{equation}
The symmetric two-rank tensor $\tilde{P}_{\mu\nu}$ is the first
trace of the variation of $E$ with respect to the Riemann tensor of
the auxiliary metric, i.e.\
\begin{equation}\label{eq:tensorP}
\tilde{P}^{\alpha\beta\mu\nu}\equiv\partial
E/\partial\tilde{R}_{\alpha\beta\mu\nu},\qquad
\tilde{P}_{\mu\nu}=\tilde{P}^\alpha_{~\mu\alpha\nu}.
\end{equation}
The second relevant definition that we will need is the
\emph{adjoint} of a differential operator $O$, denoted by $O^*$ and
which satisfies
\begin{equation}\label{eq:adjoint}
\int d^4x A\,O(B)=\int d^4x B\,O^*(A),
\end{equation}
for every pair of differential functions $A$ and $B$, with equality achieved
up to boundary terms (see \cite[p.\ 328]{Olver:1986}). The adjoint of
operator \eqref{eq:FrechetE} is then given by
\begin{equation}\label{eq:AdFrechetE}
\mathrm{D}_\mathscr{E}^*=\mathrm{D}_\mathscr{E}
-4\sqrt{-\tilde{g}}\left(\tilde{\nabla}_{\mu}\tilde{J}^{\mu}
+2\tilde{J}^{\mu}\tilde{\nabla}_{\mu}\right),
\end{equation}
with
$\tilde{J}^{\mu}\equiv\tilde{\nabla}_{\alpha}\tilde{P}^{\alpha\mu}$.

Having all the information at hand, we are now in a position
to reveal the solution to the inverse problem of the calculus of variations:
the equation $\mathscr{E}=0$ comes from an action principle iff its Fr\'echet
derivative $\mathrm{D}_\mathscr{E}$ is self-adjoint \cite[p.\
364]{Olver:1986}. According to \eqref{eq:AdFrechetE}, the operator
$\mathrm{D}_\mathscr{E}$ will be self-adjoint if for any differential
function $A$
\begin{equation}\label{eq:SelfAd}
\mathrm{D}_\mathscr{E}^*(A)=\mathrm{D}_\mathscr{E}(A) \quad
\Longleftrightarrow \quad
\tilde{J}^{\mu}=\tilde{\nabla}_{\alpha}\tilde{P}^{\alpha\mu}=0.
\end{equation}
Hence, the self-adjointness of the operator $\mathrm{D}_\mathscr{E}$
requires that the symmetric two-rank tensor \eqref{eq:tensorP} must
be conserved. Since it depends up to second order on the auxiliary
metric, we know from Lovelock theorem \cite{Lovelock:1971yv} that
the most general tensor with these properties in four dimensions is
a linear combination of the metric $\tilde{g}^{\beta\nu}$ and its
Einstein tensor $\tilde{G}^{\beta\nu}$. We finally conclude that
Helmholtz conditions \eqref{eq:SelfAd} are equivalent in four
dimensions to solve the following inhomogeneous linear first-order
PDE's
\begin{equation}\label{eq:RiemannPDE}
\tilde{g}_{\alpha\mu}\frac{\partial
E}{\partial\tilde{R}_{\alpha\beta\mu\nu}}= \lambda_1
\tilde{g}^{\beta\nu} + \lambda_2 \tilde{G}^{\beta\nu}.
\end{equation}
Actually, their solutions will completely fix the Riemann dependence of a
conformally invariant scalar equation that arises from an action principle.
This system can be rigorously integrated by splitting the dependence on the
Riemann tensor $\tilde{R}^\alpha_{~\beta\mu\nu}$ in terms of its traceless
part, the Weyl tensor $\tilde{C}^\alpha_{~\beta\mu\nu}$, the traceless part
of the Ricci tensor $\tilde{S}^\alpha_{~\beta}$, and the scalar curvature
$\tilde{R}$. It is nevertheless more instructive to exhibit the different
pieces defining the general solution. First, it is easy to show that the
variation with respect to the Weyl tensor does no contribute to the first
trace \eqref{eq:tensorP}, and hence the general solution to the homogeneous
version of Eq.~\eqref{eq:RiemannPDE}, namely $\tilde{P}^{\beta\nu}=0$, is
given by an arbitrary function of the Weyl tensor
$E_\text{h}=E_\text{h}(\tilde{g}_{\mu\nu},\tilde{C}^\alpha_{~\beta\mu\nu})$.
If this arbitrary dependence is constrained by the initial
condition $E_\text{h}(\tilde{g},0)=-4\lambda$, the coupling constant defining
the contribution of the conformal potential in \eqref{eq:Fernandestilde} is
naturally recovered. The remaining pieces can be obtained as particular
solutions of the inhomogeneous Helmholtz conditions \eqref{eq:RiemannPDE}.
The simplest one, corresponding to the part proportional to the metric, is
necessarily linear in the curvature, and can only be given by the scalar
curvature term of \eqref{eq:Fernandestilde}. Concerning the second
inhomogeneity given by the Einstein tensor, being linear in the curvature,
the particular solution must be quadratic in the curvature. The most general
quadratic combination to consider is
\begin{equation}\label{eq:squareRm}
E_\text{p}=\gamma_1\tilde{R}^2+\gamma_2\tilde{R}_{\mu\nu}\tilde{R}^{\mu\nu},
\end{equation}
since an additional contribution of the Weyl square,
$\tilde{C}_{\mu\nu\rho\sigma}\tilde{C}^{\mu\nu\rho\sigma}$, is
already taken into account as part of the homogeneous solution. A
straightforward computation yields the two-rank symmetric tensor
\begin{equation}\label{eq:Psquare}
\tilde{P}_\text{p}^{\beta\nu}=\gamma_2\tilde{G}^{\beta\nu}
+(\gamma_2+3\gamma_1)\tilde{R}\tilde{g}^{\beta\nu},
\end{equation}
which is conserved by \eqref{eq:SelfAd} only if
$\gamma_2=-3\gamma_1$. Surprisingly enough, this tuning precisely gives the
quadratic combination implementing the scalar mode suppression in the
so-called Critical Gravity \cite{Lu:2011zk}. From this term we can recover
the Gauss-Bonnet density of \eqref{eq:Fernandestilde} modulo a particular
solution of the homogenous equations, namely
\begin{equation}\label{eq:GB-W^2}
\tilde{R}^2-3\tilde{R}_{\mu\nu}\tilde{R}^{\mu\nu}=\frac32\bigl(\tilde{\mathscr{G}}
-\tilde{C}_{\alpha\beta\mu\nu}\tilde{C}^{\alpha\beta\mu\nu}\bigr).
\end{equation}
Considering all the previous results, we are able to express
the most general second-order conformally-invariant pseudoscalar arising from
an action principle for a scalar field in four dimensions as {\small
\begin{equation}\label{eq:CItildeGeneral}
\mathscr{E}=\sqrt{-\tilde{g}}\left(-\frac16\tilde{R}-4\lambda
-\alpha\tilde{\mathscr{G}} +\tilde{g}^{\alpha\beta}\frac{\partial
f}{\partial\tilde{g}^{\alpha\beta}} -2f\right)=0.
\end{equation}}%
Here, $\alpha=-3\gamma_1/2$,  and we have conveniently parameterized
the arbitrary dependence on the Weyl tensor by means of the function
$f=f(\tilde{g}_{\mu\nu},\tilde{C}^\alpha_{~\beta\mu\nu})$ which is
only constrained by our choice of initial data $f(\tilde{g},0)=0$.
Going back to the standard representation $\Phi=e^{\phi}$, the
equation (\ref{eq:CItildeGeneral}) becomes {\small
\begin{equation}\label{eq:Non-NoethCIeq}
\Box\Phi-\frac16R\Phi-\Phi^3\left(4\lambda+\alpha\tilde{\mathscr{G}}
-\tilde{g}^{\alpha\beta}\frac{\partial
f}{\partial\tilde{g}^{\alpha\beta}} +2f\right)=0,
\end{equation}}%
and describes \emph{the most general second-order Euler-Lagrange
equation for a conformal scalar field}. Of course, one could have
opted for the representation involving the Critical Gravity density
(see the conclusion) through the relation (\ref{eq:GB-W^2}) but that
would probably have hidden our main objective which was to
generalize the Fernandes conformal equation
(\ref{eq:Fernandestilde}).

Now, in order to distinguish if the new contributions have or not a
conformally Noetherian origin, we need to find the corresponding
action, whose existence is guaranteed by the generalized Helmholtz
conditions \eqref{eq:SelfAd} and \eqref{eq:RiemannPDE}. Remarkably,
this is not the end of the story since Volterra also provided an
homotopy formula to obtain the corresponding action (see \cite[p.\
364]{Olver:1986}). This was precisely the procedure employed in
Ref.~\cite{Fernandes:2021dsb}, and now, in the more general case,
it gives
\begin{align}
S[\phi,g]&=\int\!d^4x\;\phi\!\int_0^1\!ds\;\mathscr{E}|_{\phi=s\phi}
\nonumber\\
&=S_{\lambda,\alpha}[e^\phi,g] -\frac12\int
d^4x\sqrt{-\tilde{g}}f(\tilde{g}_{\mu\nu},
\tilde{C}^\alpha_{~\beta\mu\nu}). \label{eq:HomotopyAction}
\end{align}
Here, the first term corresponds to the action
\eqref{eq:actionFernandes} evaluated at $\Phi=e^\phi$, and as in the
derivation of Ref.~\cite{Fernandes:2021dsb}, we have dropped all the
$\phi$-independent pieces in the homotopy formula. It is now evident
that the new contribution is conformally invariant, and also that
the Critical Gravity density \eqref{eq:GB-W^2} is a superposition of
non-Noetherian and Noetherian conformal terms. In order to be
reassured, one can check that the variation of
\eqref{eq:HomotopyAction} with respect to the scalar field evaluated
in $\phi=\ln\Phi$ effectively yields the conformal equation
\eqref{eq:Non-NoethCIeq}. For completeness, the variation of the
action (\ref{eq:HomotopyAction}) with respect to the metric gives
the energy-momentum tensor
\begin{align}
T_{\mu\nu}={}&T^{\lambda,\alpha}_{\mu\nu}+\Phi^4\left(\frac{\partial
f}{\partial g^{\mu\nu}} -\frac12fg_{\mu\nu}\right)
+2\nabla^\rho\nabla^\sigma\left(\hat{H}_{\mu\rho\nu\sigma}\right)\nonumber\\
&+\hat{H}_{\mu\beta\rho\sigma}R_{\nu}^{~\beta\rho\sigma}
-{H}_{\mu\beta\rho\sigma}C_{\nu}^{~\beta\rho\sigma},\label{eq:Tmunu}
\end{align}
where $T^{\lambda,\alpha}_{\mu\nu}$ corresponds to the
energy-momentum tensor of action \eqref{eq:actionFernandes} as
reported in \cite{Fernandes:2021dsb}. Additionally, the four-rank
tensors stand for the Weyl variation $H^{\alpha\beta\mu\nu}=\partial
f/\partial\tilde{C}_{\alpha\beta\mu\nu}$ and its traceless part
defined by
\begin{equation}\label{eq:trlessH}
\hat{H}^{\alpha\beta}_{~~~\mu\nu}=H^{\alpha\beta}_{~~~\mu\nu}
-2\delta^{[\alpha}_{[\mu}H^{\beta]}_{\nu]}
+\frac13H\delta^{\alpha}_{[\mu}\delta^{\beta}_{\nu]},
\end{equation}
with $H^{\beta}_{\nu}=H^{\alpha\beta}_{~~~\alpha\nu}$ and
$H=H^{\beta}_{\beta}$ being its first and second traces,
respectively.

In summary, the most general non-Noetherian conformal scalar field
in four dimensions satisfying a second-order equation is explicitly
constructed thanks to the solution of the inverse problem of the
calculus of variations. We have shown that the non-Noetherian
conformal contributions are originated from a nonminimal coupling of
the scalar field to a very particular combination of
curvature-squared terms which is exactly the one defining Critical
Gravity \cite{Lu:2011zk}. In addition, we have also established that
the Noetherian conformal contributions are not restricted to the
standard ones but also include an arbitrary dependence on a
nonminimal coupling to the Weyl tensor.

This work clearly opens the door to future investigations. One of them is to
couple the full non-Noetherian conformal theory to Einstein gravity and
search for interesting solutions such as black holes, wormholes or solitons.
These solutions, if they exist, should be the natural
extension of those recently obtained in \cite{Fernandes:2021dsb}. Indeed, it
was shown in this last reference that the cohabitation between Noetherian and
non-Noetherian conformal contributions surprisingly yields to a theory which
admits black hole solutions with a phenomenological Schwarzschild-like
asymptotic. In the general case, because of the presence of the new
contribution \eqref{eq:Tmunu}, the Einstein field equations are of fourth
order which will considerably complicate their integration, but its
tracelessness can help.

The extension of this work to higher dimensions is also interesting in
physics as well as in mathematics. In fact, it is relatively immediate to
generalize our conformally invariant equation \eqref{eq:Non-NoethCIeq} to
arbitrary dimension $D$ taking into account that the conformal weight of the
scalar field \eqref{eq:ConfTransf} is now $\Phi\mapsto\Omega^{-(D-2)/2}\Phi$.
In doing so, the auxiliary metric becomes
$\tilde{g}_{\mu\nu}=\Phi^{4/(D-2)}g_{\mu\nu}$, and the generalization of the
equation \eqref{eq:Non-NoethCIeq} is given by {\small
\begin{align}
\Box\Phi-\xi_DR\Phi-\Phi^{\frac{D+2}{D-2}}\biggl[&\tfrac{2D\lambda}{D-2}
-\gamma\left(\tilde{R}^2
-\tfrac{4(D-1)}D\tilde{R}_{\mu\nu}\tilde{R}^{\mu\nu}\right)
\nonumber\\
&{}-\tilde{g}^{\alpha\beta}\frac{\partial
f}{\partial\tilde{g}^{\alpha\beta}} +\tfrac{D}2f\biggr]+\ldots=0,
\label{eq:Non-NoethCIeqD}
\end{align}}%
where $\xi_D=(D-2)/4(D-1)$ is the conformal coupling, and where the dots
stand for the higher-degree curvature contributions that Lovelock theorem
offers as possibilities in higher dimensions \cite{Lovelock:1971yv}. Here,
the Critical Gravity representation is the one that naturally appears in any
dimension \cite{Deser:2011xc}. It is particularly useful in $D=3$, where the
$\gamma$-contribution is nothing but the so-called New Massive Gravity
\cite{Bergshoeff:2009hq} and one can not make use of relations of the form
\eqref{eq:GB-W^2} since both the Gauss-Bonnet density and the Weyl tensor
identically vanish. On the other hand, the emergence of Critical Gravity
densities should not be viewed as a mystery since it can be perfectly
elucidated from the comprehensive arguments of
Refs.~\cite{Oliva:2010eb,Oliva:2010zd}. Certainly, Critical Gravities are
distinguished by the suppression of their scalar mode which
is achieved because their higher-order equations give rise to a second-order
trace. In Refs.~\cite{Oliva:2010eb,Oliva:2010zd}, Oliva and Ray concluded
that any such theory must possess a conserved second-rank tensor defined as
the first trace of the curvature variation of the Lagrangian, which is quite
similar to the tensor $\tilde{P}_{\mu\nu}$ encoding our Helmholtz conditions
\eqref{eq:SelfAd}. The authors even go further by conjecturing all the
possible forms of the curvature invariants of a given degree associated to a
conserved tensor \cite{Oliva:2010zd}. By following the lead of the Oliva-Ray
conjecture \cite{Oliva:2010zd}, one could also be able to construct the most
general second-order conformally invariant scalar equation
extremizing an action in any dimension $D$.

As a last comment, one can appreciate that equation
\eqref{eq:Non-NoethCIeqD} or its more general extension can be
viewed as a generalization of the Yamabe problem \cite{Lee:1987}.
Moreover, the fact that this equation is derivable from an action
principle makes this generalized Yamabe problem amenable to being
treated by the variational technics usually applied to the standard
version.

\begin{acknowledgments}
We would like to thank C.\ Charmousis, D.\ Flores, N.\
Lecoeur, J.\ M\'endez, J.\ Oliva, P.\ S\'anchez and M.\ San Juan for interesting
discussions and suggestions. This work has been partially
funded by Conacyt grant A1-S-11548 and FONDECYT grant 1210889.
\end{acknowledgments}

\end{document}